\documentclass[letter,twocolumn]{jpsj2} 
\usepackage{type1cm}
\usepackage[usenames]{color}

\title{%
Exact degenerate ground states \\
for the F-AF spin chain with bond alternation
}

\author{%
Hidenori \textsc{Suzuki} 
and Ken'ichi \textsc{Takano}
}

\inst{%
Toyota Technological Institute,
Tenpaku-ku, Nagoya, 468-8511
}

\recdate{\today}

\abst{%
We investigate the $J_1$-$J_2$ spin chain 
consisting of spins with magnitude $\frac12$. 
The nearest-neighbor and the next-nearest-neighbor exchange 
interactions are ferromagnetic and antiferromagnetic, 
respectively, and induce strong frustration. 
Both these interactions involve the bond alternation. 
We find exact solutions for all the degenerate ground states 
on the phase boundary of the ferromagnetic phase. 
The degeneracy remains irrespective of 
two parameters representing the bond alternation. 
The exact solutions are of closed forms for no bond alternation 
and of recursion formulae in general. 
The exact solutions are applicable to the $\Delta$ chain 
as a special case. 
}

\kword{%
quantum spin chain, frustration, ferromagnetic interaction, exact ground state
}

\newcommand{\vv}[1]{\mbox{\boldmath $#1$}}

\begin{document}	
\maketitle

The low-dimensional quantum spin system has been an interesting 
subject of numerous studies from long years ago.
In particular, the interplay of quantum fluctuation and 
geometrical frustration is of current interest.
In one-dimensional spin systems, 
various exotic phases and phenomena are reported 
such as quantum chiral 
phases \cite{nersesyan98, kaburagi99, allen00, aligia00, 
hikihara00, kolezhuk00, nishiyama00, hikihara01, kolezhuk02}, 
$\frac13$-plateau \cite{okunishi03a, okunishi03b, tonegawa04, hida05} 
and singlet cluster solid \cite{takano07}. 

The $J_1$-$J_2$ spin chain is a well-known one-dimensional 
spin model, which has the nearest-neighbor (NN) and 
the next-nearest-neighbor (NNN) interactions with 
exchange parameters $J_1$ and $J_2$, respectively. 
We restrict ourselves to the case that the spin magnitude 
$s$ is $\frac12$. 
For $J_1 > 0$ and $J_2 > 0$, 
because of the frustration, the spin chain is known to 
exhibit a quantum phase transition from a gapless 
Tomonaga-Luttinger-liquid phase 
to a gapped dimerized phase 
at $J_1/J_2 \simeq 4.15$ \cite{okamoto92,white96}.
At the Majumdar-Ghosh point of $J_1/J_2 =2$, 
the ground states are of exact tensor product forms of 
singlet NN dimers~\cite{majumdar69}.

Another type of frustration is induced 
in the $J_1$-$J_2$ spin chain, 
if the NN interaction is ferromagnetic ($J_1 < 0$) and 
the NNN interaction is antiferromagnetic ($J_2 > 0$).
We call this $J_1$-$J_2$ spin chain the F-AF chain. 
The F-AF chain is realized in, e.g., 
$\rm Rb_2 Cu_2 Mo_3 O_{12}$ 
with $J_1/J_2 \simeq -3$\cite{hase04} 
and $\rm Li Cu V O_4$ with 
$J_1/J_2 \simeq -0.3$\cite{enderle05}. 
Relatively less attention has been paid to the F-AF chain  
until such materials are discovered.  
We are interested in the difference between 
frustration effects of the F-AF chain and 
of the $J_1$-$J_2$ chain with both $J_1 > 0$ and  $J_2 > 0$.
Further, the uniform F-AF chain is extended to a model 
including bond alternation, 
if the NN and/or the NNN interactions have alternative strengths. 
The competition between frustration and bond alternation is 
of another physical interest. 

In the uniform F-AF chain, the ground state is fully ferromagnetic 
for $J_1< -4J_2$ \cite{niemeijer71,ono72,bader79}. 
For $- 4 J_2 < J_1 < 0$, numerical studies suggest 
a gapless singlet ground state\cite{white96,allen97}, 
while a detailed analysis based on the field theory predicts a tiny but non-zero 
spin-gap for small $|J_1|$\cite{itoi01}. 
At the phase boundary of $J_1 = -4J_2$, 
Hamada {\it et al.}\cite{hamada88} found the exact singlet 
ground state under the periodic boundary condition (PBC). 
The exact solution is of a resonating-valence-bond (RVB) form.  

At the phase boundary for the F-AF chain 
with the NN bond alternation, Dmitriev {\it et al.} obtained 
an exact singlet ground state 
for the PBC~\cite{dmitriev97-1,dmitriev97-2}. 
In the derivation process, they also found a special ground state 
for the open boundary condition (OBC), although it is not 
an eigenstate of the total spin. 
They further claimed that all the ground states are degenerate 
with respect to the magnitudes and the $z$-components of 
the total spin, and that the ground state for each total spin and 
each $z$-component of the total spin is unique. 
However, the explicit forms of all the degenerate ground states 
have not been shown. 

In this letter, we report all the exact degenerate 
ground states for the uniform F-AF chain under the OBC; 
they are written down in explicit forms. 
Moreover, when both the NN and the NNN bond alternations exist, 
we obtained all the exact degenerate ground states in 
simple recursion relations with respect to the system size $N$. 
The nondegeneracy of the ground state in each sector 
with the fixed total spin and its $z$-component is also shown.

\begin{figure}[tb]
\begin{center}
\includegraphics[width=0.9\linewidth]{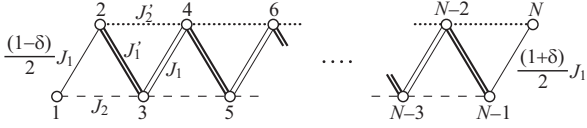}
\caption{The F-AF chain for $N$ spins with a special open 
boundary condition.
This consists of $N-2$ triangular units.}
\label{fig:chain}
\end{center}
\end{figure}

The Hamiltonian for the F-AF chain with bond alternation 
is written as 
\begin{align}
\mathcal{H} = \sum_n \, (
&J_1 \vv{s}_{2n-1}  \cdot \vv{s}_{2n}+
J_1'\vv{s}_{2n}\cdot \vv{s}_{2n+1} \nonumber\\
&+J_2 \vv{s}_{2n-1}  \cdot \vv{s}_{2n+1}+
J_2'\vv{s}_{2n}\cdot \vv{s}_{2n+2}
) , 
\label{eq:Hamiltonian} 
\end{align}
where $\vv{s}_{n}$ is the spin-$\frac12$ operator at the $n$-th site. 
$J_1$ and $J_1'$ are ferromagnetic exchange parameters 
for NN interactions, and $J_2$ and $J_2'$ are antiferromagnetic 
exchange parameters for NNN interactions. 
Different values of $J_1$ and $J_1'$ ($J_2$ and $J_2'$) 
represent the bond alternation in the NN (NNN) interactions. 
The lattice described by the Hamiltonian is illustrated 
in Fig.~\ref{fig:chain}. 
We express the NN bond alternation by $\gamma$ and 
the NNN bond alternation by $\delta$ as 
\begin{align}
J_1 = \frac{\bar{J}_1}{1 + \gamma}, &\quad 
J_1'= \frac{\bar{J}_1}{1 - \gamma}, 
\nonumber\\
J_2 = \bar{J}_2(1 - \delta) , &\quad
J_2'= \bar{J}_2(1 + \delta) , 
\label{gamma_delta} 
\end{align}
where 
$-1<\gamma<1$ and $-1\le\delta\le1$. 
Then we have the relations as 
$\bar{J}_1^{-1} = \frac{1}{2}(J_1^{-1} + {J'_1}^{-1})$ and 
$\bar{J}_2 = \frac{1}{2} (J_2 + J'_2)$. 
Owing to the parameterization 
(\ref{gamma_delta}), 
the phase boundary of the ferromagnetic phase 
is independent of $\gamma$ and $\delta$ as will be seen. 
We also define the total spin as 
$\vv{S}_{\rm tot} = \sum_{n=1}^N \vv{s}_n$, 
and denote the quantum numbers of the magnitude and 
the $z$-component by $S_{\rm tot}$ and $S_{\rm tot}^z$, 
respectively. 

Hamiltonian (\ref{eq:Hamiltonian}) is decomposed into 
triangular spin units each of which consists of three spins. 
To completely decompose it, we adopt the special OBC 
where the exchange parameter in the left end is
$J_1 (1 - \delta)/2$ and that in the right end is
$J_1 (1 + \delta)/2$ 
as shown in Fig.~\ref{fig:chain}. 
The Hamiltonian (\ref{eq:Hamiltonian}) is then rewritten as 
$\mathcal{H}=\sum_{n=1}^{N-2} \mathcal{H}_n$, 
where 
\begin{align}
\mathcal{H}_n = &
\frac{1- \delta_n}{2}  \bigg[ \bar{J}_1 
\bigg(  \frac{1}{1 - \gamma_n} \vv{s}_n\cdot \vv{s}_{n+1} \nonumber\\
&+ \frac{1}{1 + \gamma_n} \vv{s}_{n+1} \cdot \vv{s}_{n+2} \bigg)
+ 2\bar{J}_2  \vv{s}_n\cdot \vv{s}_{n+2} \bigg ] 
\label{eq:H_triangle} 
\end{align}
with $\delta_n \equiv (-1)^{n} \delta$ and 
$\gamma_n \equiv (-1)^{n} \gamma$. 

Solving the Hamiltonian $\mathcal{H}_n$ 
for a single triangular unit, 
all the different eigenvalues divided by $1- \delta_n$ are 
\begin{align}
\frac{\bar{J}_1+\left(1-\gamma ^2\right) \bar{J}_2}{4 \left(1-\gamma ^2\right)} , 
\quad 
-\frac{\bar{J}_1+\left(1-\gamma ^2\right) \bar{J}_2 
\pm \sqrt{D}}
{4 \left(1-\gamma ^2\right)} 
\label{tri_eigenvalues} 
\end{align}
with $D = 3 \gamma ^2 \bar{J}_1^2+\left(\bar{J}_1-2 \left(1-\gamma ^2\right) \bar{J}_2\right)^2$. 
The composite spin of the three spins is $\frac{3}{2}$ 
for the first and $\frac{1}{2}$ for the second 
in eq.~(\ref{tri_eigenvalues}). 
For $\bar{J}_1< -4\bar{J}_2$, the first one 
in eq.~(\ref{tri_eigenvalues}) is the lowest. 
We can incorporate the spin-$\frac{3}{2}$ states 
of all the triangular units into a ferromagnetic 
state of the total Hamiltonian $\mathcal{H}$. 
This means that the ferromagnetic state is 
the ground state of $\mathcal{H}$. 
Similar argument has been done in the case of 
no bond alternation~\cite{bader79}. 
For $\bar{J}_1 > -4\bar{J}_2$, 
the lowest eigenvalue for a triangular unit is the second one 
with the upper sign in eq.~(\ref{tri_eigenvalues}). 
We cannot incorporate the spin-$\frac{1}{2}$ states 
of all the triangular units into an eigenstate of $\mathcal{H}$.
In particular, the ground state is not ferromagnetic 
for $\bar{J}_1 > -4\bar{J}_2$. 
Thus the condition $\bar{J}_1$ = $-4\bar{J}_2$ represents 
the phase boundary of the ferromagnetic phase 
irrespective of the values of $\gamma$ and $\delta$. 

In the uniform case of $\gamma = \delta = 0$, 
we found all the exact degenerate ground states 
in explicit forms at the phase boundary of $J_1 = -4J_2$.
The exact ground state with 
$S_{\rm tot}=S_{\rm tot}^z=\frac{N}{2}-p$ 
is written down as 
\begin{align}
& |\Phi_p \rangle = c_p \sum_{\{i_k, j_k\}} 
|\phi_p(i_1,j_1;i_2,j_2;\ldots;i_p,j_p)\rangle ,
\label{eq:Phi_p} 
\\
& |\phi_p(i_1,j_1;i_2,j_2;\ldots;i_p,j_p)\rangle = 
\prod_{q=1}^{p} | (i_q, j_q) \rangle 
\prod_{l\neq \{i_k, j_k\}}|\uparrow_{l}\rangle,
\label{eq:phi_p} 
\end{align}
where $c_p$ is a normalization constant, 
and $| (i_q, j_q) \rangle \equiv$
$(|\uparrow_{i_q}\downarrow_{j_q}\rangle - 
|\downarrow_{i_q} \uparrow_{j_q} \rangle ) / \sqrt2$ 
is the singlet pair of spins at $i_q$ and $j_q$. 
Here, $|\uparrow_i\rangle$ and $|\downarrow_i\rangle$ for the single spin 
labeled by $i$ are eigenstates of $s_i^z$ with 
eigenvalues $\frac12$ and $-\frac12$, respectively.
The summation in eq.~(\ref{eq:Phi_p}) has been taken 
over all possible combinations 
$\{i_1,j_1;i_2,j_2;\ldots;i_p,j_p\}$, or shortly $\{i_k, j_k\}$, 
under the condition that $i_k < j_k$ $(k = 1, 2, \cdots , p)$. 
Clearly the state (\ref{eq:Phi_p}) with $p=0$ is 
fully ferromagnetic.
In the special case of $p=\frac{N}{2}$ with even $N$, 
the state (\ref{eq:Phi_p}) is the same as the exact singlet ground 
state derived by Hamada {\it et al.}\cite{hamada88}.
By operating $\vv{s}_{n_1}\cdot \vv{s}_{n_2}$ 
on $|\phi_p(i_1,j_1;\cdots)\rangle$ in eq.~(\ref{eq:Phi_p}), 
we find 
\begin{align}
& \vv{s}_{n_1}\cdot \vv{s}_{n_2}|\phi_p (i_1,j_1;\cdots)\rangle 
	= \frac14|\phi_p (i_1,j_1;\cdots)\rangle, \\
& \vv{s}_{n_1}\cdot \vv{s}_{n_2}|\phi_p (n_1,j_1;\cdots) \rangle  
\nonumber\\
&{} \quad = \frac14 |\phi_p (n_1,j_1;\cdots)\rangle 
	-\frac12|\phi_p (n_1,n_2;\cdots)\rangle, \\
& \vv{s}_{n_1}\cdot \vv{s}_{n_2}|\phi_p (n_1,n_2;\cdots)\rangle 
	= -\frac34|\phi_p (n_1,n_2;\cdots)\rangle , \\
&\vv{s}_{n_1}\cdot \vv{s}_{n_2}
\big[|\phi_p(n_1,i_1;n_2,j_2;\cdots)\rangle+  |\phi_p(n_1,j_2;n_2,i_1;\cdots)\rangle \big] 
\nonumber\\
&{} \quad = \frac14 \big[ |\phi_p(n_1,i_1;n_2,j_2;\cdots)\rangle 
	+ |\phi_p(n_1,j_2;n_2,i_1;\cdots)\rangle \big] , 
\end{align} 
where $i_k$ and $j_k$ $(k=1, 2, \cdots , p)$ are different 
from $n_1$ or $n_2$. 
By the aid of these equations, we obtain 
\begin{align}
\mathcal{H}_n |\Phi_p\rangle = -\frac{3 J_2}{4}  |\Phi_p\rangle
\label{eq:eigHn} 
\end{align}
for all $n$. 
Summing up this equation from $n=1$ to $N-2$, 
we have the following eigenvalue equation: 
\begin{align}
\mathcal{H} |\Phi_p\rangle = -\frac{3J_2}{4}(N-2)  |\Phi_p\rangle.
\label{eq:eigH} 
\end{align}
This eigenvalue is the ground state energy, 
because it is just the lower bound which is the sum of 
the lowest energies of the $N-2$ triangular units  
as known from eq.~(\ref{tri_eigenvalues}) with 
$\gamma = \delta = 0$ and $J_1 = -4J_2$. 
Therefore $|\Phi_p\rangle$ for any $p$ is an exact ground state.

To find the ground state for arbitrary $S_{\rm tot}$ and 
$S_{\rm tot}^z$, we tilt $\vv{S}_{\rm tot}$ 
by symmetrically flipping some of half spins in 
the ferromagnetic part of $|\phi_p\rangle$.
Then we have 
\begin{align}
|\Phi_{pp'}\rangle = c_{pp'} & \sum_{\{i_k,j_k\}} 
\Bigg[ \prod_{q=1}^{p} | (i_q, j_q) 
\rangle  \nonumber \\ 
& \times
{\sum}'\Bigg(\prod_{l=1}^{p'} |\downarrow_{l}\rangle
\prod_{l'\neq \{ i_k,j_k,l \}} |\uparrow_{l'}\rangle \Bigg)\Bigg],
\label{eq:Phipp'} 
\end{align}
where $c_{pp'}$ is the normalization constant and 
the primed summation is taken over all $_{N-2p}C_{p'}$ 
combinations of $p'$ down spins and
$N-2p-p'$ up spins for each $\{i_k,j_k\}$.
This ground state $|\Phi_{pp'}\rangle$ has $S_{\rm tot}=\frac{N}{2}-p$ 
and $S_{\rm tot}^z=\frac{N}{2}-p-p'$.
Equation (\ref{eq:eigH}) and 
the spin rotational symmetry of $\mathcal{H}$ guarantee that 
$|\Phi_{pp'}\rangle$ is a ground state. 
The number of states in the form of eq.~(\ref{eq:Phipp'}) 
is $\frac14 (N+2)^2$ for even $N$ and 
$\frac14 [(N+2)^2-1]$ for odd $N$.

For the general F-AF chain with bond alternation, 
it is difficult to write down all the degenerate 
ground states in explicit forms. 
We however found that the ground states are exactly expressed 
in simple recursion relations with respect to the system size $N$ on the phase boundary of $\bar{J}_1=-4\bar{J}_2$. 
The recursion relations are derived from the fact that 
a ground state of the total chain is simultaneously the ground 
states of $\mathcal{H}_n$ for all the triangular units. 
The derivation is explained in what follows. 

Let $|j,m \rangle_N$ be the ground state of the $N$-site 
chain for $S_{\rm tot}=j$ and 
$S_{\rm tot}^z=m$ at $\bar{J_1}=-4\bar{J}_2$. 
The $(N+2)$-sites ground state $|j,m \rangle_{N+2}$ is expressed 
in terms of the $N$-sites ground states for 
$S_{\rm tot}=j\pm1$ and $j$ with states of 
two extra $\frac12$ spins at 
the $(N+1)$-th and the $(N+2)$-th sites. 
By denoting the Clebsch-Gordan coefficient as 
${\rm C}(\nu,\mu) = \langle 1,\mu ; j+\nu,m-\mu | j,m \rangle$, 
it is written as follows: 
\begin{align}
|j,m\rangle_{N+2} =
& a_{N}(j) \sum_{\mu} 
{\rm C}(1,\mu)
| j+1,m-\mu \rangle_{N} \otimes |{\rm t}_\mu \rangle \nonumber\\
 + & b_{N}(j) \sum_{\mu} 
{\rm C}(0,\mu)
| j,m-\mu \rangle_{N} \otimes |{\rm t}_\mu \rangle\nonumber\\
 + & c_{N}(j) \sum_{\mu} 
{\rm C}(-1,\mu)
| j-1,m-\mu \rangle_{N} \otimes |{\rm t}_\mu \rangle\nonumber\\
 + & d_{N}(j) | j,m-\mu \rangle_{N} \otimes |{\rm s} \rangle,
\label{eq:N} 
\end{align}
where $|{\rm s}\rangle$ is the singlet state for the extra spins, 
and $|{\rm t}_\mu\rangle$ is the triplet state for them 
with quantum number  $\mu$ of 
the $z$-component of the composite spin; then 
each summation takes over $\mu = -1$, 0 and 1. 
The coefficients $\{a_{N}, b_{N}, c_{N}, d_{N}\}$ are 
independent of $m$ because of 
the rotational symmetry of $\mathcal{H}$. 
Since $|\frac{N}{2}+1,m \rangle_{N+2}$ is fully ferromagnetic, 
we immediately find $a_N(\frac{N}{2}+1) = 0$, 
$b_N(\frac{N}{2}+1) = 0$, $c_N(\frac{N}{2}+1) = 1$ and $d_N(\frac{N}{2}+1) = 0$. 
Since there do not exist $|\frac{N}{2}+1,m \rangle_{N}$ and 
$|j-1,m \rangle_{N}$ for $j\le\frac12$, we have 
$a_{N}(\frac{N}{2})=0$ and $c_N(0) = c_N(\frac12) = 0$. 
Further, since $|0,0\rangle_{N+2}$ cannot be produced from
$|0,0\rangle_{N}$ and $|{\rm t}_\mu \rangle$, we have 
$b_N(0)=0$. 

Using eq.~(\ref{eq:N}) two times successively, we obtain 
$|j,m \rangle_{N+2}$ in the following form:
\begin{align}
|j,m \rangle_{N+2} = \sum_{\{s_i^z=\pm 1/2\}} 
&|A_N(s_{N-1}^z, s_{N}^z, s_{N+1}^z, s_{N+2}^z)\rangle 
\nonumber \\
&\otimes 
|s_{N-1}^z s_{N}^z s_{N+1}^z s_{N+2}^z \rangle.
\label{eq:fourspins} 
\end{align}
Here, $|A_N(s_{N-1}^z, s_{N}^z, s_{N+1}^z, s_{N+2}^z)\rangle$ 
is expressed by a summation of the ground states of $(N-2)$-site 
chain and contains $\{a_{N}, b_{N}, c_{N}, d_{N}\}$ 
and  $\{a_{N-2}, b_{N-2}, c_{N-2}, d_{N-2}\}$.
The recursion relations for $\{ a_N, b_N, c_N, d_N \}$ 
are derived by imposing the condition that eq.~(\ref{eq:fourspins})
is the ground state of the local Hamiltonian $\mathcal{H}_{N-1}$ 
and $\mathcal{H}_N$. 
Then we have 
\begin{align}
(\mathcal{H}_{N-1}+\mathcal{H}_N)|j,m \rangle_{N+2}
=-\frac{(3+\gamma^2) \bar{J}_2}{2(1-\gamma^2)} \, 
|j,m \rangle_{N+2}.
\label{eq:eigH_N} 
\end{align}
This eigenvalue equation with eq.~(\ref{eq:fourspins}) stands, 
only if the following recursion relations are satisfied: 
\begin{align}
& \frac{a_{N}(j)}{d_{N}(j)} 
	= \frac{a_{N-2}(j)}{d_{N-2}(j+1)}, 
\label{eq:rec_a}\\
& \frac{b_{N}(j)}{d_{N}(j)} 
	= \frac{b_{N-2}(j)}{d_{N-2}(j)}, 
\label{eq:rec_b}\\
& \frac{c_{N}(j)}{d_{N}(j)} 
	= \frac{c_{N-2}(j)}{d_{N-2}(j-1)} .
\label{eq:rec_c} 
\end{align}
Further, the following relations for the same $N$ should be satisfied: 
\begin{align}
&\frac{b_N(j)}{d_N(j)}
= \sqrt{\frac{j-1}{j+1}} \, \frac{b_N(j-1)}{d_N(j-1)}
+\frac{4}{1+\gamma}\sqrt{\frac{j}{j+1}},
\label{eq:b_N(j)}
\\
&\frac{c_N(j)}{d_N(j)} 
= \left[\sqrt{j-1} b_N(j-1) + \sqrt{j} d_N(j-1)\right] \nonumber\\ 
&\times\left[\sqrt{j-1} b_N(j-1) + \left(\frac{3-\gamma}{1+\gamma}\right) \sqrt{j} d_N(j-1)\right]\nonumber\\ 
&\times\left[\sqrt{(2j+1)(2j-1)} a_N(j-1)d_N(j-1) \right]^{-1}.
\label{eq:c_N(j)} 
\end{align}
With the normalization condition
\begin{align}
a_{N}^2+b_{N}^2+c_{N}^2+d_{N}^2 = 1, 
\label{eq:normaliz} 
\end{align}
the recursion relations (\ref{eq:rec_a}) to (\ref{eq:rec_c}) 
determine $\{a_{N}, b_{N}, c_{N}, d_{N}\}$ 
by starting from initial values for, e.g., $N=4$ 
except for coefficients with special values of $j$. 
Since the denominator of the right hand side of 
eq.~(\ref{eq:rec_a}) for $j=\frac{N}{2}-1$ and 
of eq.~(\ref{eq:rec_b}) for $j=\frac{N}{2}$ becomes zero, 
we separately evaluate $a_{N}(\frac{N}{2}-1)$ and $b_{N}(\frac{N}{2})$ 
by using eqs.~(\ref{eq:b_N(j)}) and (\ref{eq:c_N(j)}). 
Parameter $\gamma$ representing the NN bond alternation 
only affects eqs.~(\ref{eq:b_N(j)}) and (\ref{eq:c_N(j)}). 
All the above equations are independent of $\delta$.

\begin{table}[tbp]
\begin{center}
\begin{tabular}{c|cccc}
$j$ & 0 & 1 & 2 & 3\\
\hline
$a_4(j)$ 
	& $ (3 - \gamma)\alpha_0 \beta_0 $ 
	& $ (5 - \gamma) (3+\gamma ) \sqrt{2}\, \alpha_1 $ 
	& 0 
	& 0 \\
$b_4(j)$ 
	& 0 
	& $ 2 \sqrt{30} \, \alpha_1 \beta_0 $
	& $ 2 \sqrt{6} \, \alpha_2 $
	& 0 \\
$c_4(j)$ 
	& 0 
	& $ \sqrt{5}\, \alpha_1 \beta_1 $
	& $ \sqrt{2}\, \alpha_2 \beta_0 $
	& 1 \\
$d_4(j)$ 
	& $ \alpha_0 \beta_1 $ 
	& $ (1+\gamma) \sqrt{15} \, \alpha_1 \beta_0 $
	& $ (1+\gamma)\alpha_2 $
	& 0
\end{tabular}
\caption{Initial coefficients $\{a_4, b_4, c_4, d_4\}$ 
for the recursion relations (\ref{eq:rec_a}) to (\ref{eq:rec_c}).
Here, $\alpha_0 = (51+10\gamma^2+3\gamma^4)^{-1/2}$, 
$\alpha_1 = \{3(385+200\gamma+86\gamma^2+24\gamma^3+9\gamma^4)\}^{-1/2}$, 
$\alpha_2 = (35+6\gamma+3\gamma^2 )^{-1/2} $, 
$\beta_0 = (5+2\gamma+\gamma^2)^{1/2} $ and $\beta_1 = (1+\gamma)\{2(3+\gamma^2)\}^{1/2}$.}
\label{tab:N=4}
\end{center}
\end{table}

The initial coefficients $\{a_4, b_4, c_4, d_4\}$ 
for the recursion relations (\ref{eq:rec_a}) to (\ref{eq:rec_c}) 
are given in Table~\ref{tab:N=4}, and the initial ground states 
$\{|j,m\rangle_4\}$ are uniquely determined as 
\begin{align}
&|2,2\rangle_4 
	= |\uparrow\uparrow\uparrow\uparrow\rangle, 
\qquad |2,1\rangle_4 
	= \frac{(1,1,1,1) \cdot \vv{u}_1}{2}, \nonumber \\
& |1,1\rangle_4
	= \frac{(3+\gamma ,1-\gamma, -1+\gamma, -3-\gamma) \cdot \vv{u}_1}
		{2\sqrt{5+2\gamma+\gamma^2}}, \nonumber \\
& |2,0\rangle_4
	= \frac{(1,1,1,1,1,1) \cdot \vv{u}_0}{\sqrt6}, \nonumber \\
& |1,0\rangle_4
	= \frac{(2,1+\gamma,0,0,-1+\gamma,-2) \cdot \vv{u}_0}{\sqrt{2(5+2\gamma +\gamma^2 )}}, \nonumber \\
& |0,0\rangle_4
	= \frac{(3-\gamma,2\gamma,-3+\gamma,-3+\gamma,2\gamma,3+\gamma ) \cdot \vv{u}_0}
		{2\sqrt{3(3+\gamma^2 )}}, \nonumber \\
& |2,-1\rangle_4
	= \frac{(1,1,1,1) \cdot \vv{u}_{-1}}{2}, \nonumber \\
& |1,-1\rangle_4
	= \frac{(3+\gamma ,1-\gamma, -1+\gamma, -3-\gamma) \cdot \vv{u}_{-1}}
		{2\sqrt{5+2\gamma +\gamma^2}}, \nonumber \\
& |2,-2\rangle_4 
 	= |\downarrow\downarrow\downarrow\downarrow\rangle, 
\label{eq:gs_4}
\end{align} 
where 
$\vv{u}_1 \equiv ( |\uparrow\uparrow\uparrow\downarrow\rangle 
, |\uparrow\uparrow\downarrow\uparrow\rangle
, |\uparrow\downarrow\uparrow\uparrow\rangle 
, |\downarrow\uparrow\uparrow\uparrow\rangle)^{\rm t}$, 
$\vv{u}_0\equiv(|\uparrow\uparrow\downarrow\downarrow\rangle 
, |\uparrow\downarrow\uparrow\downarrow\rangle
, |\downarrow\uparrow\uparrow\downarrow\rangle)
, |\uparrow\downarrow\downarrow\uparrow\rangle 
, |\downarrow\uparrow\downarrow\uparrow\rangle 
, |\downarrow\downarrow\uparrow\uparrow\rangle)^{\rm t}$ and 
$\vv{u}_{-1}\equiv( |\uparrow\downarrow\downarrow\downarrow\rangle
, |\downarrow\uparrow\downarrow\downarrow\rangle
, |\downarrow\downarrow\uparrow\downarrow\rangle
, |\downarrow\downarrow\downarrow\uparrow\rangle)^{\rm t}$.
Thus the ground state $|j,m\rangle_N $ 
with arbitrary even $N$ is constructed recursively
by eqs. (\ref{eq:N}), (\ref{eq:rec_a}) to (\ref{eq:c_N(j)}). 
Note that all the coefficients can be determined 
as positive values by a suitable choice of the sign of each ground state.
Thus the coefficients $\{a_{N}(j), b_{N}(j), c_{N}(j), d_{N}(j)\}$ are determined uniquely. 
Further, since the total ground state is also a ground state of each 
local Hamiltonian $\mathcal{H}_n$, any total state which is not a local ground 
state cannot be another total ground state. 
Therefore, the ground state in the sector of 
fixed $S_{\rm tot}=j$ and $S_{\rm tot}^z=m$ is 
nondegenerate. 
Although we have only shown the initial values for even $N$, 
the above recursion relations 
is valid for odd $N$ by using the initial coefficients and the ground states for $N=3$. 
The total degeneracy of the ground states is 
$\frac14 (N+2)^2$ for even $N$ and 
$\frac14 [(N+2)^2-1]$ for odd $N$.
Among the ground states for the OBC 
the state for $j=m=0$ is simultaneously the 
ground state for the PBC 
even if $\gamma\neq0$ and $\delta \neq 0$. 

\begin{figure}[tb]
\begin{center}
\includegraphics[width=1.0\linewidth]{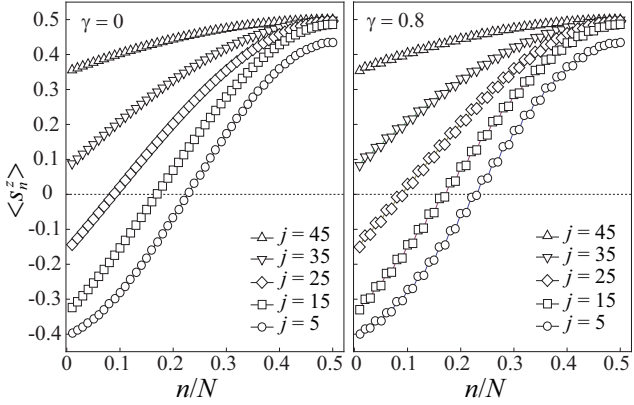}
\caption{The expectation value $\langle S_n^z \rangle$ 
for the ground state $|j,j\rangle_{100}$ 
when $\gamma=0$ (left) and $\gamma =0.8$ (right).}
\label{fig:snz}
\end{center}
\end{figure}

Using the above recursion relations, 
we can calculate physical quantities for large $N$.
In Fig.~\ref{fig:snz}, we show the expectation value 
$\langle S_n^z \rangle$ for the ground state $|j,j\rangle_{100}$ 
when $\gamma=0$ and $\gamma=0.8$.

We now examine the $\Delta$ chain, which is represented 
by Hamiltonian $\mathcal{H}$ with $J_2' = 0$ ($\delta = -1$) 
and odd $N$. 
Since $\mathcal{H}_n = 0$ for even $n$, 
eq.~(\ref{eq:eigH_N}) reduces to 
\begin{align}
\mathcal{H}_N|j,m \rangle_{N+2}
=-\frac{(3+\gamma^2) \bar{J}_2 }{2(1-\gamma^2)} |j,m \rangle_{N+2}, 
\end{align} 
which is independent of $s_{N-1}$. 
Then, instead of eqs.~(\ref{eq:rec_a}) to (\ref{eq:c_N(j)}), 
we have 
\begin{align}
& \frac{a_{N}(j)}{a_{N-2}(j)} = 
	\frac{\sqrt{j} b_{N}(j) + \left(\frac{3-\gamma}{1+\gamma} \right)
	\sqrt{j+1} d_{N}(j)}{\sqrt{j+2} b_{N-2}(j+1) - \sqrt{j+1} d_{N-2}(j+1) },\\
& \frac{c_{N}(j)}{c_{N-2}(j)} = 
	\frac{\sqrt{j+1} b_{N}(j) - \left(\frac{3-\gamma}{1+\gamma} \right)
	\sqrt{j} d_{N}(j) }{\sqrt{j-1} b_{N-2}(j-1) + \sqrt{j} d_{N-2}(j-1)}, \\
& \frac{\sqrt{2 j+1} c_N(j) }{\sqrt{j+1} b_N(j) - \sqrt{j} d_N(j)}\nonumber\\
	&\qquad\qquad =
	\frac{\sqrt{j-1} b_N(j-1) + \sqrt{j} d_N(j-1) }{\sqrt{2 j-1} a_N(j-1)}. 
\label{eq:parameter_delta} 
\end{align}
These equations are not enough to uniquely determine 
the coefficients $\{a_N,b_N,c_N,d_N\}$, 
since the degeneracy of the ground states increases. 
We speculate the $_{(N-1)/2}C_{j-1/2}$-fold degeneracy 
for given $N$, $j$ and $m$ by numerical calculations. 

To summarize, we found exact solutions of 
all the degenerated ground states of the F-AF chain 
on the ferromagnetic phase boundary. 
For a uniform F-AF chain, 
the exact ground states for arbitrary $S_{\rm tot}$ and 
$S_{\rm tot}^z$ is written down explicitly in the closed form 
(\ref{eq:Phipp'}).
In each state, $p$ singlet pairs are distributed uniformly 
among the remaining $N-2p$ unpaired spins.
For a general F-AF chain with bond alternation, 
the recursion formulae for the ground states 
with respect to the system size $N$ have been derived. 
This formulae is independent of the NNN bond alternation. 
By using the recursion formulae, 
we can evaluate various physical quantities of the ground states 
for large systems. 
The detailed calculations will be reported elsewhere.

This work is partly supported by 
Fund for Project Research in Toyota Technological Institute.


\end{document}